\documentclass[preprint,showpacs,floats,floatfix,nofootinbib]{revtex4}

\usepackage{amsbsy}
\usepackage{amssymb}
\usepackage{amsmath}
\input{epsf}
\usepackage{graphicx}
\usepackage{amssymb}
\usepackage{bm}
\def\spose#1{\hbox to 0pt{#1\hss}}

\def\lta{\mathrel{\spose{\lower 3pt\hbox{$\mathchar"218$}}
     \raise 2.0pt\hbox{$\mathchar"13C$}}}
\def\gta{\mathrel{\spose{\lower 3pt\hbox{$\mathchar"218$}}
     \raise 2.0pt\hbox{$\mathchar"13E$}}}

\usepackage{color}

\def\beq{\begin{equation}}
\def\eeq{\end{equation}}
\def\bea{\begin{eqnarray}}
\def\eea{\end{eqnarray}}

\def\n{{\rm n}}

\def\n{{\rm n}}

\begin{document}
\title{A variational approach to relativistic superfluid vortex elasticity}

\author{N. Andersson$^1$, S. Wells$^1$ and G.L. Comer$^2$  }
\affiliation{$^1$\ Mathematical Sciences and STAG Research Centre, University of Southampton, Southampton 
SO17 1BJ, UK \\
$^2$\ Department of Physics, Saint Louis University, St. Louis, MO 63156-0907, USA}

\date{\today}

\begin{abstract}
It is well known that a superfluid rotates by forming an array of quantized vortices. A relativistic formulation for superfluid vortex dynamics is required for a range of problems in astrophysics and cosmology, from neutron star interiors and radio pulsar glitches to possible dark matter condensates on galactic scales. This paper develops a formalism for  such systems, extending the well-established variational approach to relativistic fluids to account for the presence of a collection of quantized vortices. The model is firmly anchored in the geometry of the problem (drawing on aspects from basic string dynamics) and accounts for elastic aspects associated with a vortex array, providing a precise foundation for applications which have so far been based on somewhat ad hoc phenomenology.
\end{abstract}

\maketitle

\section{Introduction}

Superfluids mimic bulk rotation by forming an array of quantized vortices---a collection of slim ``tornadoes'', the distribution of which determines the macroscopic angular momentum of the system.
The associated dynamics plays a key role in the description of superfluid systems, both in the laboratory setting and in astrophysics \cite{sfbook,sfns}. In particular, an understanding of vortex dynamics is thought to be essential for an explanation of the enigmatic spin glitches seen in many young pulsars. The presence of a large-scale neutron superfluid in both the core and the crust (where is co-exits with a lattice of neutron-rich nuclei) is a pre-requisite for any realistic  description of a mature neutron star, and the simple fact that these systems involve extreme densities (reaching several times the nuclear saturation density) means that the relevant modelling has to be done in the context of relativistic gravity. Given this, there have been efforts to extend models for relativistic fluid dynamics to account for superfluid components and the presence of quantized vortices. A recent example---with close connection to the discussion in this paper---provided the first description of the vortex-mediated mutual friction \cite{stuart}, a dissipation channel that is unique to superfluids and which is known to be important for models of macroscopic neutron star dynamics \cite{men,trevmf}. 

The fact that vortices are associated with a long-range interaction---which is how they contribute to bulk rotation---implies that the vortex lattice has elastic properties \cite{baymc,cbaym,baym}. This is an interesting aspect, which may be of observational/experimental relevance. In particular, the vortex lattice supports a set of elastic oscillation modes. These so-called Tkachenko modes, first proposed in a seminal set of papers in the 1960s \cite{tk1,tk2}, have been discussed for superfluid helium, superfluid atomic condensates \cite{bec1,bec2,bec3} and neutron stars \cite{ns1,ns2,ns3,ns4}. The experimental verification of the idea is, however, quite recent \cite{cod}. The main difficulty is that other aspects of the physics (vortex pinning to the surface of a laboratory container, or the neutron star crust, and the mixing with inertial modes of a rotating body) tend to overwhelm the subtle effect of the vortex elasticity \cite{sonin}.

Despite the experimental issues, the problem  of vortex elasticity is of obvious conceptual interest. Nevertheless, the problem has neither been considered within general relativity, nor within the approach of ``modern'' elasticity theory (where the elastic properties are viewed as due to the deviation from a relaxed/unstrained equilibrium configuration \cite{cq1,cq2,KS}). With this paper, we aim to fill this gap in the discussion. Drawing on the variational approach to relativistic fluid dynamics  \cite{carter,livrev}, and basic aspects of string dynamics \cite{letel,stachel,vile}, we provide a  description of vortex elasticity (notably based on a two-dimensional subspace, orthogonal to the vortex array).  As this description is---at least in principle---fully nonlinear, it goes beyond previous (more phenomenological) discussions of the problem \cite{baymc,cbaym,baym}, which tend to be rooted in perturbation theory. Moreover, by considering our final results from the perturbative point of view, we shed light on the origin of expressions that have been used in applications and indicate how these models can be extended to the curved spacetime setting.

\section{Variational fluid model}

In order to set the scene, and provide both context and inspiration for the discussion, it makes sense to review the standard variational approach to relativistic fluids (following the approach from  \cite{carter} and \cite{livrev}). It may seem somewhat odd to go over supposedly familiar material in detail, but it turns out to be relevant to compare and contrast the derivation of the fluid equations with the strategy for the vortex lattice.  In essence, the fluid derivation makes use of a three-dimensional matter space in order to ensure that the matter flux $n^a$ is conserved (thus constraining the variation). In the vortex case, we will introduce an analogous---now two-dimensional---subspace, in order to enforce constraints on the vorticity.  

Let us first consider a single matter component, represented by a (conserved) flux $n^a$ (with spacetime indices $a,b,c,\ldots = \{0,1,2,3\}$). 
For an isotropic system the matter Lagrangian, which we will call $\Lambda$, should be a relativistic invariant and hence 
depend only on $n^2 = -g_{ab} n^a n^b$. In effect, this means that the Lagrangian depends on the flux and the space-time metric. 
An arbitrary variation of $\Lambda=\Lambda(n^2)=\Lambda(n^a,g_{ab})$ then gives (ignoring terms that can be written as total derivatives, representing``surface terms'' in the action) 
\beq
    \delta \left(\sqrt{- g} \Lambda\right) = \sqrt{- g} \left[\
    \mu_a \delta n^a + \frac{1}{2} \left(\Lambda g^{a b} +  
    n^a \mu^b\right) \delta g_{a b}\right] \ , \label{dlamb}
\eeq
where $g$ is the determinant of the spacetime metric and $\mu_a$ is the canonical  momentum:
\beq
     \mu_{a} = {\partial \Lambda \over \partial n^a} = -2 {\partial \Lambda \over \partial n^2} g_{ab} n^b \ .
     \label{momdef}
                   \end{equation} 
We have also used
\begin{equation}
\delta \sqrt{-g} =  {1\over 2} g^{ab} \delta g_{ab}  \ .
\end{equation}

Equation \eqref{dlamb} illustrates why we  need to develop a constrained variational principle. As it stands, the variation of $\Lambda$ suggests that the equations of motion would be $\mu_a=0$, which means that the fluid carries neither  energy nor momentum. This problem is resolved by constraining the flux. A natural way to do this involves introducing a three-dimensional ``matter space'', the coordinates of which,
$X^A$ with $A,B,C,\ldots = \{1,2,3\}$, serve as  labels that distinguish fluid element worldlines, assigned at the initial time of the evolution, say $t=0$. 
The matter space coordinates can be considered as scalar fields on spacetime, with a unique map (obtained by a pull-back construction) relating them to the spacetime coordinates:
\begin{equation}
\psi^A_a = {\partial X^A\over \partial x^a} \ .
\end{equation} 
With this set-up, the conservation of the matter flux is ensured provided that the dual  three-form
\begin{equation}
 n_{abc} = \epsilon_{abcd} n^{d} = - n^d  \epsilon_{dabc} \quad , \quad
    n^a = \frac{1}{3!} \epsilon^{a b c d} n_{b c d}  \ , \label{n3form}
 \end{equation}
 (where $\epsilon_{abcd}$ is the volume form associated with spacetime) is closed. It is easy to see that
\begin{equation} 
     \nabla_{[a} n_{bcd]} = 0\quad   \Longrightarrow \quad 
  \nabla_{a} n^{a} = 0 \ . \label{consv2} 
\end{equation} 

Let us consider this argument in more detail.
The closure is guaranteed by introducing
\beq
    n_{a b c} =  \psi^A_{[a}\psi^B_b\psi^C_{c]}
                   n_{A B C} \ , \label{pb3form}
\eeq 
where the (anti-symmetric) matter-space $n_{ABC}$ depends only on the $X^A$ coordinates (and the Einstein summation convention applies to repeated matter-space indices). In order for this to make sense, $n_{abc}$ must be  a ``fixed'' tensor, in the sense that\footnote{Note that the four-velocity $u^a$ is associated with the fluid flow in this instance. This will be different when we focus on the vorticity later.}
\beq
u^a n_{abc} = 0  \ ,
\eeq
and
\beq
\mathcal L_u n_{abc} = 0 \ .
\label{Lien}
\eeq
The latter is equivalent to requiring
\beq
\nabla_{[a}n_{bcd]} = \partial_{[a}n_{bcd]} = 0\ .
\eeq
The final step involves noting that 
\beq
\partial_{[a}n_{bcd]} = \psi^A_a \psi^B_b \psi^C_c\psi^D_d \partial_{[A} n_{BCD]} = 0 \ ,
\eeq
is automatically satisfied if 
\beq
\partial_{[A} n_{BCD]} = 0 \ ,
\eeq
which, in turn, follows if $n_{ABC}$ is a function only of the $X^A$ coordinates. This completes the argument.

Formally, we have changed perspective by taking the (scalar fields) $X^A$   to be fundamental 
variables.  
The construction also provides matter space with a geometric structure. 
If integrated over a volume in matter space, $n_{ABC}$ provides a measure of the number of particles in that volume. In essence, we have
\begin{equation}
n_{ABC} = n\epsilon_{ABC} \ .
\end{equation}

The final step in the derivation of the fluid equations involves introducing the 
Lagrangian displacement $\xi^a$, tracking the motion of a given fluid element. From the standard definition
of  Lagrangian variations,  we have 
\beq
\Delta X^A = \delta X^A + \mathcal{L}_{\xi} X^A = 0 \ , 
\label{DelX}
\eeq
where $\delta X^A$ is the Eulerian variation and $\mathcal{L}_{\xi}$ is the Lie derivative along $\xi^a$. This means that we have
\beq
    \delta X^A = -  \mathcal{L}_{\xi} X^A = -  \xi^a {\partial X^A \over \partial x^a} = -\xi^a \psi^A_a\ . 
    \label{xlagfl}
\end{equation} 
It also follows that
\beq
\Delta \psi^A_a = 0 \ ,
\eeq
and
\beq
\mathcal L_u \psi^A_a =   0\ .
\eeq
Given these results, it is easy to show that 
\beq
\Delta n_{abc} = 0 \ ,
\eeq
a fact that will be useful later.

Making use of the standard relations
\beq
\delta g_{db} =  - g_{da} g_{bc} \delta g^{ac} \ ,
\eeq
and
\beq
 \delta \epsilon^{abcd} = {1\over 2}  \epsilon^{abcd} g_{ef} \delta g^{ef}  \ ,
\eeq
we now have
\beq
\delta n^a  = {1\over 3!} \delta ( \epsilon^{a b c d} n_{b c d} )  = n^b \nabla_b \xi^a - \xi^b 
                   \nabla_b n^a - n^a \left(\nabla_b 
                   \xi^b -  \frac{1}{2} g_{b c} \delta 
                   g^{b c}\right) \ .
                     \label{delnvec} 
\end{equation} 
Expressing the variations of the matter Lagrangian in terms of the displacement $\xi^a$, rather than the perturbed flux, we 
ensure that the flux conservation is accounted for in the equations of motion. 
The variation of 
$\Lambda$ then leads to 
\beq
    \delta \left(\sqrt{- g} \Lambda\right) = \sqrt{- g} \left\{  f_a \xi^a -  \frac{1}{2}\left[\left( \Lambda - n^c\mu_c\right) g_{a b} +  
    n_a \mu_b \right] \delta g^{a b}  \right\}  \ , \label{variable}
\eeq
and the fluid equations of motion are given by
\begin{equation} 
     f_b \equiv 2 n^a \nabla_{[a} \mu_{b]}   = 0 \ ,
     \label{force}
\eeq
 (where the square brackets indicate anti-symmetrization, as usual). Finally, introducing the vorticity two-form
 \beq
 \omega_{ab} = 2\nabla_{[a} \mu_{b]}  \ ,
 \label{omdef}
 \eeq
 we have the simple relation
 \beq
 n^a \omega_{ab} = 0 \ .
 \eeq
 
 We can also read off the stress-energy tensor from \eqref{variable}. We need 
\begin{equation}
T_{ab} = - {2 \over \sqrt{-g}} {\delta \left( \sqrt{-g}\Lambda\right) \over \delta g^{ab}}  =  \Lambda g_{ab} - 2 {\delta \Lambda \over \delta g^{ab}} \ .
\end{equation}
Introducing the matter four-velocity, such that $n^a=nu^a$ and $\mu_a = \mu u_a$, where $\mu$ is the chemical potential, we see that the energy is
\begin{equation}
\varepsilon = u_a u_b T^{ab} = - \Lambda \ .
\end{equation}
Moreover, we identify the pressure from the thermodynamic relation:
\begin{equation}
p = -\varepsilon + n\mu  =  \Lambda - n^c\mu_c  \ .
\end{equation}
This means that we have
\begin{equation}
T^{ab} = pg^{ab} + n^a \mu^b = \varepsilon u^a u^b + p h^{ab} \ ,
\label{stressen0}
\end{equation}
where we have used the standard projection
\begin{equation}
h^{ab} = g^{ab} + u^a u^b  \ .
\end{equation}

Finally, it is straightforward to confirm that
\begin{equation} 
\nabla_a T^{ab} = - f^b + \nabla^b \Lambda - \mu^b \nabla_a n^a = - f^b = 0 \ ,
\label{divT}
\end{equation}
since i) $\Lambda$ is a function only of $n^a$ and $g_{ab}$, and ii) the definition of the momentum $\mu_a$.

Up to this point we have rehearsed standard arguments, but  it turns out that it pays off to keep the detailed steps in mind as we proceed. 

\section{The Kalb-Ramond approach}

In order to make cautious progress, we now set out to derive the fluid results from a different perspective. The ultimate aim is to arrive at an intuitive description of  the (suitably averaged) dynamics of a collection of quantized superfluid vortices. 

The strategy builds on efforts to relate  string dynamics to the forces acting on a superfluid vortex,  first considered in \cite{lund,kalb} and developed further in \cite{ps1,ps2}. 
We start by noting that the superfluid fluid velocity (technically, the momentum \cite{livrev}) can be linked  the gradient of a scalar potential $\alpha$ such that $\tilde H_a = \partial_a \alpha$.
The key idea is to identify this velocity as the dual\footnote{From this point  on, we use tildes to indicate Hodge duals.} 
 \beq
 \tilde H_a = {1\over 3!} \epsilon_{abcd} H^{bcd} \ ,
 \eeq
and introduce the so-called Kalb-Ramond field \cite{kalb}, such that
 \beq
 H^{abc} =  \nabla^{[a}B^{bc]}=  \partial^{[a} B^{bc]} \ .
 \eeq
 It is now easy to see that the scalar wave equation
 \beq
\Box \alpha = 0 \ ,
 \eeq 
 is automatically satisfied, as long as
 \beq
 \nabla_a \left( \nabla^a B^{bc} + \nabla^c B^{ab} + \nabla^b B^{ca} \right) = 0 \ .
 \label{bwave}
 \eeq
 In effect, we can shift the focus from $\alpha$ to $B^{ab}$. Alternatively, we could treat $B^{ab}$ as an independent field (and try to solve the more complicated wave equation \eqref{bwave}). The relevant dynamical equations are then automatically solved by expressing this field in terms of a scalar potential---the two descriptions are complementary \cite{ps1}. The advantage of the  Kalb-Ramond representation may not be particularly clear at this point, but we will soon see that it makes the introduction of topological defects (vortices/strings) intuitive.

As a first step in this direction, we return to the fluid problem but shift the attention from the matter flux to the vorticity. 
Following \cite{kr1,kr2,kr3}, we do this by noting that we can ensure that the conservation law \eqref{consv2} is automatically satisfied by introducing a two-form $B_{ab}$  (the Kalb-Ramond field) such that
\beq
n_{abc} = 3 \nabla_{[a}B_{bc]} \ .
\label{Bdef}\eeq
That is, we have
\beq
n^a = {1\over 2} \epsilon^{abcd} \nabla_b B_{cd} \ ,
\eeq
and the flux conservation \eqref{consv2}  follows as an identity\footnote{By construction $n_{abc}$ is exact, which means that it is automatically closed.}---we no longer need to introduce the three-dimensional matter space. 

Noext, in order to find an action that reproduces the known perfect fluid results,  we elevate 
 the vorticity $\omega_{ab}$ to an additional variable. A Legendre transformation \cite{kr2} leads to the Lagrangian\footnote{The motivation for the transformation is that we take the vorticity $\omega_{ab}$ to be the conjugate variable associated with the flux $n^a$.  As the vorticity is defined in terms of the momentum $\mu_a$, which would be the ``usual'' conjugate, this is not every different from the traditional approach.  Moreover, as we will see later, the specific form of \eqref{newlambda} is chosen to reproduce the fluid result (see \eqref{vortdef}) and leaves the  stress-energy tensor unaffected.} 
\beq
\bar \Lambda = \Lambda - {1\over 4} \epsilon^{abcd} B_{ab} \omega_{cd}  = \Lambda - {1\over 2} \tilde \omega^{ab} B_{ab} \ ,
\label{newlambda}
\eeq
where we have used the dual
\beq
\tilde \omega^{ab} = {1\over 2}  \epsilon^{abcd} \omega_{cd} \ .
\eeq

Assuming that $\Lambda =\Lambda(n)$ we get (ignoring the perturbed metric for the moment)
\beq
\delta \bar \Lambda = -{1\over 3!} \mu^{abc} \delta n_{abc} - {1\over 2} B_{ab} \delta \tilde \omega^{ab} - {1\over 2} \tilde \omega^{ab} \delta B_{ab} \ ,
\eeq
where we have introduced 
\beq
 {\partial \Lambda \over n_{abc} } = - {1\over 3!} \mu^{abc} \ .
\eeq
However, we also have
\beq
\delta n_{abc} = 3 \nabla_{[a} \delta B_{bc]} \ ,
\eeq
which means that
\beq
\delta \bar \Lambda  ={1\over 2}  \left(  \nabla_{a}  \mu^{abc} -  \tilde \omega^{bc}  \right) \delta B_{bc}- {1\over 2} B_{ab} \delta \tilde \omega^{ab} -{1\over 2} \nabla_a \left( \mu^{abc} \delta B_{bc}\right) \ .
\label{krvar}\eeq
Ignoring the surface term (as usual), we see that a variation with respect to $B_{ab}$ requires
\beq
\tilde \omega^{bc} =  \nabla_{a}  \mu^{abc} \ ,
\label{vortdef}\eeq
which leads us back to \eqref{omdef}.
However, with a free variation we would also have $B_{ab}=0$.  With the free lunch as elusive as ever, we need to constrain the variation of $\tilde\omega^{ab}$ (or rather $\omega_{ab}$). Fortunately, the matter space argument from the original fluid derivation provides us with the strategy for doing this. 

In order for the vorticity to be a purely spatial object---orthogonal to the flow--we must have\footnote{At this point, we need to appreciate the difference between fluid elements and topological defects like vortices/strings.The former are naturally associated with worldlines, the tangent vector of which provides the four-velocity $u^a$. In contrast, a vortex is associated with a two-dimensional world sheet. This world sheet is spanned by two vectors, one timelike and one spacelike. In our discussion, we take the timelike vector be the four velocity. This is an important distinction because it means that $u^a$ is less directly linked to the motion of the ``fluid'', which (still) follows from $n^a$. These notions should become clear as we progress.}
\beq
u^a\omega_{ab} = 0  \ .
\eeq
In addition, we want is to be ``fixed'' in the (new) matter space, in the sense that
\beq
\mathcal L_u \omega_{ab} = 0 \ .
\eeq
Since $\omega_{ab}$ is anti-symmetric, this leads to 
\beq
u^c \nabla_{[a}\omega_{bc]} = 0 \ .
\eeq
Clearly, this condition will be satisfied if
\beq
\nabla_{[a}\omega_{bc]} = \partial_{[a}\omega_{bc]} = 0  \ .
\label{two}\eeq

The key difference is that we now make use of a two-dimensional space with coordinates $\chi^I$ (here, and in the following $I,J,\ldots$ represent two-dimensional coordinates). We  obtain this two-dimensional space either via a map from the original matter space
\beq
\hat \psi^I_A = {\partial \chi^I \over \partial X^A} \ ,
\eeq
or directly from spacetime, using
\beq
\bar \psi^I_a = {\partial \chi^I \over \partial x^a}\ .
\eeq
The two descriptions are (obviously) consistent since
\beq
\bar \psi^I_a =\hat \psi^I_A  \psi^A_a  = {\partial \chi^I \over \partial X^A}  {\partial X^A \over \partial x^a} = {\partial \chi^I \over \partial x^a} \ ,
\eeq
via the chain rule. The coordinates and the corresponding maps are illustrated in Figure~\ref{maps}.

\begin{figure}
\includegraphics[width=0.8\textwidth]{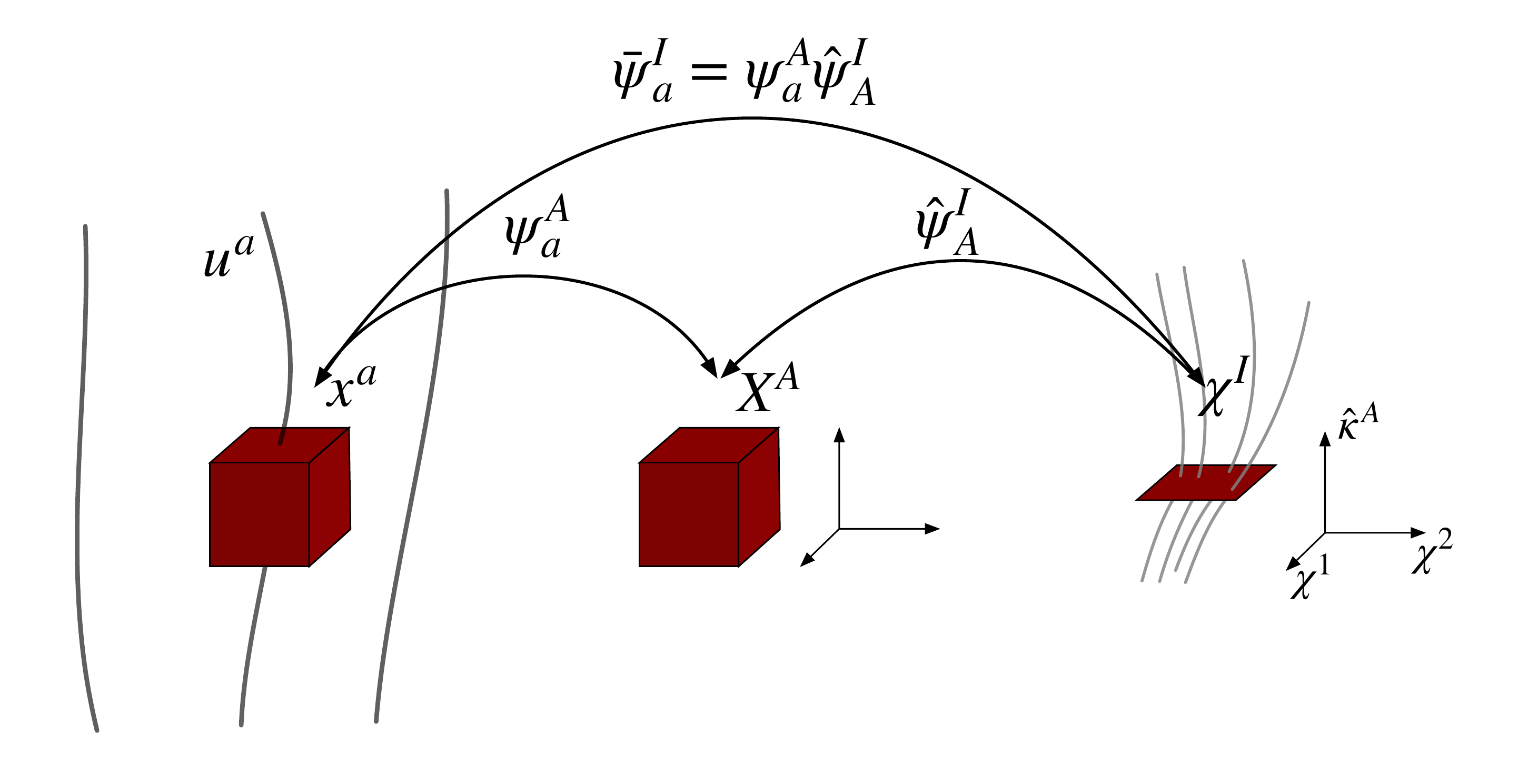}
\caption{An illustration of the different coordinates and maps used in the discussion of vortex dynamics and elasticity.}
\label{maps}
\end{figure}

 apting the logic that led to the conserved matter flux, we  introduce the matter space tensor $\omega_{IJ}$, such that 
\beq
\omega_{ab} = \psi^A_a \psi^B_b \omega_{AB} = \bar \psi^I_a \bar \psi^J_b \omega_{IJ} \ .
\eeq
Noting that \eqref{two} becomes
\beq
 \partial_{[a}\omega_{bc]} = \bar \psi^I_a \bar \psi^J_b \bar \psi^K_c \partial_{[I} \omega_{JK]} = 0 \ ,
 \label{vortder}
\eeq
it follows that the required condition holds provided  $\omega_{IJ}$ only depends on the $\chi^I$ coordinates. The logic is quite familiar.

Next, with
\beq
\Delta \chi^I = 0 \quad  \Longrightarrow \quad \delta \chi^I = - \mathcal L_\xi \chi^I \ ,
\eeq
 we  have 
\beq
\Delta \omega_{ab} = 0 \ ,
\eeq
which, again ignoring the metric perturbations, leads to
\beq
\delta \tilde \omega^{ab} = {1\over 2} \epsilon^{abcd} \delta \omega_{cd}
= - \xi^c \nabla_c \tilde \omega^{ab} - \epsilon^{abcd} \omega_{ed} \nabla_c \xi^e \ .
\eeq
After some algebra, we find that the middle term in \eqref{krvar} leads to (leaving out surface terms)
\beq
- {1\over 2} B_{ab} \delta \tilde \omega^{ab} 
 = {3\over 2} \xi^c \tilde \omega^{ab} \nabla_{[c} B_{ab]} \ ,
\eeq
where have have noted that \eqref{vortdef} implies the conservation law 
\beq
\nabla_a \tilde \omega^{ab} = 0 \ .
\label{vcons}
\eeq
We now see that a variation with respect to $\xi^a$ leads to 
\beq
{3\over 2}  \tilde \omega^{ab} \nabla_{[c} B_{ab]} = {1\over 4}  \epsilon^{abde} \omega_{de}   n_{cab}  =   n^d \omega_{dc} = 0 \ ,
\eeq
and we recover the usual  fluid equations of motion.

We still do not seem to have made much progress, but the introduction of a two-dimensional ``vortex space'' is essential if we want to explore the dynamics of a collection of quantized vortices. This will become clear shortly.   

For convenience, let us also derive the stress-energy tensor within the new ``strategy''. 
Taking \eqref{newlambda} as our starting point and noting that
\beq
n^2  = - n^a n_a = - {1\over (3!)^2} g_{ah} \epsilon^{abcd} \epsilon^{hefg} n_{bcd} n_{efg} \ .
\eeq
We have
\beq
\delta \Lambda = {\partial \Lambda \over \partial n^2} \delta n^2 
 = -  {2\over 3!}  {\partial \Lambda \over \partial n^2} \left( n_d  \epsilon^{dabc}\right)  \delta n_{abc}
 +  {\partial \Lambda \over \partial n^2}   \left( n_a n_b  + n^2  g_{ab}  \right)\delta g^{ab} \ .
 \label{varlam}
\eeq
We also need 
\beq
- {1\over 4} \delta \epsilon^{abcd} B_{ab} \omega_{cd}  =
- {1\over 4} \tilde \omega^{cd} B_{cd}  g_{ab} \delta g^{ab}  \ ,
\eeq
to get
\beq
{\delta {\bar \Lambda} \over \delta g^{ab} } =   {\partial \Lambda \over \partial n^2}   \left( n_a n_b  + n^2  g_{ab}  \right) - {1\over 4} \tilde \omega^{cd} B_{cd}  g_{ab} \ .
\eeq
Finally, 
\beq
T_{ab} = \bar \Lambda g_{ab} - 2 {\delta \bar \Lambda \over \delta g^{ab}}= \Lambda  g_{ab} - 2  {\partial \Lambda \over \partial n^2}   \left( n_a n_b  + n^2  g_{ab}  \right)  \ ,
\label{Ttest1}
\eeq 
leads us back to \eqref{stressen0} once we recall the definition of the momentum \eqref{momdef}. 

This completes the argument. The introduction of the Kalb-Ramond field shifts the focus onto the vorticity,  which is naturally associated with a two-dimensional subspace (replacing the usual three-dimensional matter space). The key point is that we arrive at the fluid equations without explicitly associating the fluid flux $n^a$ with the four-velocity $u^a$. Let us now consider this point in more detail.

\section{String interlude}

In order to form a complete picture---including connections with related problems---and develop some of the tools we need to make progress, it is (perhaps not surprisingly) natural to take a detour in the direction of string theory. 

A one-dimensional string moving through spacetime traces out a two-dimensional world sheet\footnote{The world sheet aspect is common between the string problem and that of vortex dynamics. The key difference is that strings tend to be taken to move through a vacuum, whereas a vortex lives in a medium---typically a superfluid condensate. The geometric aspects of the two problems are close, even though some of the physics aspects are different.}. This world sheet is naturally spanned by two vectors, one timelike (intuitively taken to be the four velocity of the string $u^a$) and one spacelike (naturally, the tangent vector to the string, represented by $\hat \kappa^a$). These vectors are associated with two-dimensional coordinates\footnote{When combined, the two sets of coordinates $\phi^I$ and $\chi^I$ provide us with the means the represent spacetime.} such that  $x^a = x^a (\phi^I)$, leading to
 the tangent surface element
 \beq
S^{ab} = \epsilon^{IJ} {\partial x^a \over \partial \phi^I} {\phi x^b \over \partial \phi^J} \ ,
\label{surf}
 \eeq
 with $\epsilon^{IJ}$ the (normalised) two-dimensional Levi-Civita tensor (density).

Associated with this world sheet we have a bivector (read: an anti-symmetric tensor of rank 2), let us call is $\Sigma^{ab}$, mathematically parameterised in terms of the two linearly independent vectors that span the surface (as the bivector represents a surface, it is natural to think of it as a contravariant object). Noting that   a simple timelike bivector can be written as the alternating product of a timelike and a spacelike vector \cite{stachel} (such that its dual will be a simple spacelike bivector) and  assuming the normalisation 
\beq
 \Sigma_{ab} \Sigma^{ab} = -2  \ ,
 \label{norm}
 \eeq
we  use
\beq
 \Sigma^{ab} = u^a  \hat\kappa^b  - u^b \hat \kappa^a \ ,
 \label{biv}
\eeq
such that
\beq
\hat \kappa^a = \Sigma^{ab} u_b \ .
\eeq
The projection into the two-dimensional space spanned by $u^a$ and $\hat \kappa^a$ is then given by
 \beq
 \Sigma^{ac} \Sigma_{cb} =  \hat \kappa^a \hat \kappa_b - u^a u_b  \ .
 \eeq
 
 Introducing the dual
\beq
\tilde \Sigma_{ab} = {1\over 2} \epsilon_{abcd} \Sigma^{cd}  =   \epsilon_{abcd} u^c \hat\kappa^d \ ,
\eeq
we also have the  orthogonal projection
 \beq
 \perp^a_{\ b} = \tilde \Sigma^{ac} \tilde \Sigma_ {cb} = \delta^a_b + u^a u_b - \hat \kappa^a \hat \kappa_b \ ,
 \eeq
 and it follows that
\beq
\tilde \Sigma_{ab} \Sigma^{bc} = 0 
\eeq
Note, for later convenience, that this results follows immediately from the condition that the bivector is simple:
\beq
\Sigma^{[ab} \Sigma^{c]d} = 0 \quad \Leftrightarrow \quad \Sigma^{ab} \Sigma^{cd} \epsilon_{abce}  = 0 \ .
\label{simple}
\eeq
Finally, the bivector is surface forming, provided that \cite{stachel}
 \beq
\tilde \Sigma_{ab} \nabla_c \Sigma^{bc} =  \tilde \Sigma_{ab} \partial_c \Sigma^{bc} = 0  \ .
 \label{surfo}
 \eeq
  
With this set-up, we may take the bivector to be proportional to the surface element. Letting
\beq
\Sigma^{ab} = \alpha^{-1/2} S^{ab}  \ ,
\eeq
we have
\beq
\Sigma^{IJ} = \alpha^{-1/2} S^{IJ} = \alpha^{-1/2}  \epsilon^{IJ} \ .
\eeq
Making use of the  induced metric (which we also use to raise and lower indices in the two-dimensional subspace)
 \beq
 \gamma_{IJ} = g_{ab} {\partial x^a \over \partial \phi^I} {\partial x^b \over \partial \phi^J} \ ,
 \eeq
 we have
 \beq
  \gamma_{IK} \gamma_{JL}  \epsilon^{IJ}  \epsilon^{KL} = - 2 \alpha \ ,
 \eeq
 and we identify
 \beq
 \alpha = - \gamma = - \mathrm{det}\ \gamma_{AB}\ .
 \eeq
 That is, we have
 \beq
 \Sigma^{ab} = \sqrt{-\gamma} {\partial x^a \over \partial \phi^I} {\partial x^b \over \partial \phi^J}  \epsilon^{IJ} \ .
 \eeq
 Geometrically, the dual of $\Sigma^{ab}$ is a two-form that represents (when integrated) the flux carried by vortices (strings) across a surface. The variable $\gamma$ is a measure of this flux.

 It is now natural to assume\footnote{The essence of the argument is that we want to find the metric that minimizes the area. In two dimensions all metrics are conformally flat, so there is only one degree of freedom to vary, and it is natural to associate this with the determinant $\gamma$.} that the Lagrangian of the system depends on $\gamma$, with 
 \beq
 \gamma =  {1\over 2} \Sigma^{ab} \Sigma_{ab} = {1\over 2} \Sigma^{IJ} \Sigma_{IJ} \quad (=-1) \ .
 \eeq
 Moreover, as we want to compare and contrast with a model based on averaging over a network of vortices---essentially treated as a fluid described by a small number of fields (density, velocity, tension etcetera)---it is natural to consider the example of a coarse-grained ``string fluid'' \cite{shub1,shub2,shub3}. Hence, we take 
 $\sqrt{-g} \Lambda(\gamma)$ to be the matter contribution to the action, noting that, if we let $\Lambda = M \sqrt{-\gamma}$ this leads to the coarse-grained version of the standard Nambu-Goto string action \cite{letel,vile}, with $M$ the string tension.

For the stress-energy tensor we need
\beq
\delta \Lambda = {d\Lambda \over d\gamma} \left( {\partial \gamma \over \partial \Sigma^{ab} } \delta \Sigma^{ab} +  {\partial \gamma \over \partial g_{ab}} \delta g_{ab} \right) 
+ {d\Lambda \over d\gamma} \left( \Sigma_{ab} \delta \Sigma^{ab} + \Sigma_c^{\ a} \Sigma^{cb} \delta g_{ab} \right) \ ,
\eeq
so
\beq
T^{ab}  =  \Lambda g^{ab} + 2 {\delta \Lambda \over \delta g_{ab}} = \Lambda g^{ab} + 2 {d\Lambda \over d\gamma} \Sigma_c^{\ a} \Sigma^{cb} \ .
\eeq
From this it follows that the equations of motion are
\beq
\nabla_a T^{ab} = g^{ab} \nabla_a \Lambda + 2  \Sigma_c^{\ a} \Sigma^{cb} \nabla_a \left( {d\Lambda \over d\gamma} \right) +  2 {d\Lambda \over d\gamma} \nabla_a   \left( \Sigma^a_{\ c}  \Sigma^{cb}\right) = 0 \ .
\eeq
 However, since $\gamma=-1$ we have $\nabla_a \gamma=0$, which means that we only need
  \begin{multline}
 \nabla_a   \left( \Sigma^a_{\ c}  \Sigma^{cb}\right) 
 =  \Sigma^{cb} \nabla_a \Sigma^a_{\ c}  + {1 \over 2} \Sigma_{ca} \left(  \nabla^a  \Sigma^{cb} +\nabla^c \Sigma^{ba} + \nabla^b \Sigma^{ca} \right) 
 \\
 =  \Sigma^{cb} \nabla_a \Sigma^a_{\ c}  + 3 \Sigma_{ca}   \nabla^{[a}  \Sigma^{cb]}  =0 \ ,
  \end{multline}
  where we have used \eqref{norm}.
Following \cite{stachel}, we contract with $\Sigma_{db}$ to get 
  \beq
 \Sigma_{db}  \Sigma^{cb} \nabla_a \Sigma^a_{\ c}  + 3 \Sigma_{[ac} \Sigma_{b]d}   \nabla^{[a}  \Sigma^{cb]}  =0 \ ,
  \eeq
  where the second term vanishes since the bivector is simple, cf. \eqref{simple}. Noting also that 
  \beq
   \Sigma_{db}  \Sigma^{cb} \nabla_a \Sigma^a_{\ c} = 0 \quad \Longrightarrow \quad \Sigma_{dc} \nabla_a \Sigma^{ac} =0 \ ,
  \eeq
  and considering \eqref{surfo}, we infer the conservation law \cite{stachel,shub3} 
  \beq
  \nabla_a \Sigma^{ab} = 0 \ .
  \label{conse}
  \eeq
  In essence, if the contractions of a vector with both the bivector and its dual vanish then the vector must be zero.  Returning to the equations of motion, we are left with 
 \beq
  \Sigma^a_{\ c}  \nabla_a  \Sigma^{cb} = 0 \ ,
  \eeq
  or
  \beq
   \perp^c_b \left( \hat \kappa^a \nabla_a \hat \kappa^b - u^a \nabla_a u^b \right) = 0  \ .
  \eeq
This is the simplest version of the model---describing how the surface tension serves to drive the system towards a minimum area---which will be sufficient for our purposes. Still, it is interesting to note extensions like the dissipative case considered in \cite{shub3} and the discussion of charged cosmic strings in \cite{cart}.

 Before we move on, let us establish two useful results. First of all, we have
  \beq
\hat \kappa^a = \Sigma^{ab} u_b \Longrightarrow  \nabla_a \hat \kappa^a  + u^a u_b  \nabla_a \hat \kappa^b = u_b \nabla_a \Sigma^{ab} = 0 \ ,
\label{divk}
 \eeq
 by virtue of \eqref{conse}. 
 Similarly 
 \beq
  u^a = \Sigma^{ab} \hat \kappa_b   \Longrightarrow
  \nabla_a u^a  - \hat \kappa^a \hat \kappa_b  \nabla_a u^b = \hat \kappa_b \nabla_a \Sigma^{ab} = 0 \ .
  \label{divu}
 \eeq

\section{Vortex dynamics}

A natural extension to the model developed in Section~III allows $\Lambda$ to depend on both $n_{abc}$ and $\omega_{ab}$ from the outset. Intuitively,  such a model represents a superfluid condensate with (averaged) vorticity represented by a collection of vortices. At the quantum level, the dynamics would be represented by a single wave function, but at the fluid level we can always describe the problem in terms of an irrotational condensate and a contribution from vortices. 

Starting from $\Lambda = \Lambda (n_{abc}, \omega_{ab}, g^{ab})$ we immediately have 
(using the convention from \cite{kr2})
\beq
\delta \Lambda = -{1\over 3!} \mu^{abc} \delta n_{abc} - {1\over 2} \lambda^{ab} \delta \omega_{ab} + {\delta \Lambda \over \delta g^{ab}} \delta g^{ab} \ ,
\label{dlambf}\eeq
where
\beq
\lambda^{ab} = - 2 {\partial \Lambda \over \partial \omega_{ab} } \ .
\eeq
The first term in \eqref{dlambf} may be interpreted as the energy cost associated with introducing additional particles in the system, while the second term is associated with rotational energy. 
From \eqref{newlambda} it then follows that (ignoring the metric variation and a surface term, as before)
\beq
\delta \tilde \Lambda 
= {1\over 2} \left( \nabla_{c}  \mu^{cab}- \tilde \omega^{ab} \right) \delta B_{ab}  - {1\over 2} \left(  \lambda^{cd} + {1\over 2}  \epsilon^{abcd} B_{ab} \right) \delta \omega_{cd} \ ,
\eeq
which leads us back to \eqref{krvar} and  \eqref{vortdef}. However,  we now have an additional term involving $\delta \omega_{ab}$. 
Making use of \eqref{two}, this new term can be written
\beq
-{1\over 2}   \lambda^{cd}  \delta \omega_{cd} = {1\over 2}  \lambda^{cd}\left( \xi^a \nabla_a \omega_{cd} + 2 \omega_{ad} \nabla_c \xi^a \right) \\
=  - \xi^a  \omega_{ad} \nabla_c \lambda^{cd} \ .
\eeq
Combining this  with the result from the previous section, we see that a variation with respect to the displacement leads to   (see \cite{kr1,kr2,kr3})
\beq
n^a \omega_{ab} =  \omega_{ab} \nabla_c \lambda^{ca} = - 2  \omega_{ab} \nabla_c \left({\partial \Lambda \over \partial \omega_{ca} } \right) \ .
\label{neweuler}
\eeq
Basically, the explicit dependence on the vorticity has led to amended equations of motion. However, if we want to interpret the term on the right-hand side of \eqref{neweuler} we need to do a little bit more work. 

First of all, it is worth noting that we may write \eqref{neweuler} as
\beq
\left[ n^a + 2  \nabla_c \left({\partial \Lambda \over \partial \omega_{ca} } \right) \right] \omega_{ab}  \equiv \bar n^a \omega_{ab} = 0 \ ,
\eeq
with 
\beq
\bar n^a = n^a + 2  \nabla_c \left({\partial \Lambda \over \partial \omega_{ca} } \right)\ .
\eeq
This means that $\bar n^a$ must be proportional to $u^a$, which
 makes the result appear more ``familiar'' (see Section~II) but it does not really help us understand the  ingredients in \eqref{neweuler}. 

Instead, let us  consider the implications of the two-dimensional matter space. Intuitively, the idea makes sense for a collection of (locally) aligned quantized vortices as one can always introduce a two-dimensional surface orthogonal to the vortex array (that is, orthogonal to the world sheet we used in the discussion of strings). Points in this surface are described by the $\chi^I$ coordinates. 
Not surprisingly, we can adapt the logic from the usual matter-space construction to this new setting---although  in doing so we would focus on the map from the original three-dimensional space to the two-dimensional one. As is evident from \eqref{vortder}, we  also need the map from spacetime to either of the two lower-dimensional spaces. In essence, the original fluid derivation involved
\beq
\psi^A_b \psi^a_A = h^a_{\ b} = \delta^a_b+u^a u_b\ .
\eeq
Meanwhile, the corresponding map to the two-dimensional stage takes the form 
\beq
\hat \psi^I_B \hat \psi^A_I = \delta^A_B - \hat \kappa^A \hat \kappa_B\ ,
\eeq
with a suitable spatial unit vector $\hat \kappa^a$, automatically orthogonal to the four velocity $u^a$ since
\beq
u^a \hat \kappa_a = (u^a \psi^A_a) \hat \kappa^A = 0 \ .
\eeq
We  take the new vector $\hat \kappa^a$ to be normal to the area spanned by the $\chi^I$ coordinates (and identify it with the spacelike coordinate we used in the discussion of the string world sheet). That is, we have
\beq
\hat \kappa^A \hat \psi^I_A = 0 \ .
\eeq
In essence, $\hat \kappa^A$ is assumed to be aligned with the quantized vortices. 
It also follows that 
\beq
\bar \psi^I_a \bar \psi^b_I = ( \psi^A_a \hat \psi^I_A)  ( \psi^b_B \hat \psi^B_I)  = \psi^A_a \psi^b_B (\delta_A^B - \hat \kappa_A \hat \kappa^B ) 
=  \delta_a^b+u_a u^b - \hat \kappa_a \hat \kappa^b \equiv  \perp^b_a \ .
\eeq
This will be relevant later.

In order to stress the close resemblance to the various relations for $n_{ABC}$ from section~II,  we first of all introduce a vector
\beq
W^A = {1\over 2} \epsilon^{ABC} \omega_{BC} \quad \Longrightarrow  \quad \omega_{AB} = \epsilon_{ABC} W^C \ .
\eeq
In spacetime, we then have
\beq
W^a =  {1\over 2} \psi^a_A \epsilon^{ABC} \omega_{BC} =  {1\over 2} \psi^a_A \psi^b_B \psi^c_C \epsilon^{ABC} \omega_{bc} = {1\over 2} u_d \epsilon^{dabc} \omega_{bc}\ .
\eeq
We recognize this as the vorticity  vector \cite{stuart} and note that it is simply related to the dual of the vorticity;
\beq
W^a = u_d \tilde \omega^{da}\ .
\eeq

We may also work in the two-dimensional space, where we must have 
\beq
\omega_{IJ} = \mathcal N \kappa \epsilon_{IJ} \quad \Longrightarrow  \quad\omega_{AB} = \mathcal N \kappa  \epsilon_{AB} \ ,
\eeq
with (for future reference)
\beq
\epsilon_{IJ} \epsilon^{JK} = \delta_I^K \ ,
\eeq
\beq
\epsilon_{IJ} \epsilon^{IJ} = 2\ ,
\eeq
and
\beq
\epsilon_{AB} = \hat \kappa^C \epsilon_{CAB}\ .
\eeq	

Letting $\kappa^A = \kappa \hat \kappa^A$, we now have
\beq
 \omega_{AB} = \mathcal N  \kappa^C \epsilon_{CAB} \ ,
\eeq
so
\beq
\kappa^A \omega_{AB} = 0 \ .
\eeq
In fact, we have
\beq
W^A = \mathcal N \kappa^A\ .
\eeq
The interpretation of this is intuitive---we have a collection of vortices, each associated  with a quantum $\kappa$ of circulation---with number density (per unit area) $\mathcal N$.

We also have
\beq
W^2 = (\mathcal N \kappa)^2 =   {1\over 2} \omega_{IJ} \omega^{IJ} = {1\over 2} \omega_{AB} \omega^{AB} =  {1\over 2} \omega_{ab} \omega^{ab} 
= {1\over 2} g^{ac} g^{bd} \omega_{ab} \omega_{cd} \ .
\eeq

Finally, the spacetime vorticity takes the (expected)  form 
\beq
\omega_{ab} = \mathcal N u^c \kappa^d \epsilon_{cdab}\ ,
\label{omnew}
\eeq
(explicitly connecting to the dual $\tilde \Sigma_{ab}$ used to describe string dynamics in Section~IV).
We also have
\beq
\mathcal L_u \kappa_a = \mathcal L_u \left( \psi_a^A \kappa_A \right) = \psi_a^A  \mathcal L_u \kappa_A  = \psi^A_a u^c \partial_c \kappa_A
=  \psi^A_a (u^c  \psi^B_c) {\partial \kappa_A\over \partial X^B } = 0 \ ,
\eeq
\beq
u^b \nabla_b  \mathcal N = (u^b \tilde \psi^I_b) {\partial \mathcal N \over \partial \chi^I} = 0\ ,
\label{uNperp}
\eeq 
and
\beq
\kappa^a \nabla_a \mathcal N = \kappa^a \tilde \psi^I_a  {\partial \mathcal N \over \partial \chi^I}  = \psi_A^a \kappa^A \psi^B_a \hat \psi^I_B   {\partial \mathcal N \over \partial \chi^I}  = \kappa^A \delta^A_B \hat \psi^I_B  {\partial \mathcal N \over \partial \chi^I}  = \kappa^A \hat \psi^I_A  {\partial \mathcal N \over \partial \chi^I}  = 0 \ .
\label{kNperp}
\eeq	
These results are quite intuitive, and (for later convenience) it is worth noting that 
\beq
    (u^a u_b - \hat \kappa^a \hat \kappa_b )  \nabla_a \mathcal N  = 0 \ ,
    \label{Nperp2}
 \eeq
 and we will also need to recall \eqref{divk} and \eqref{divu}. 
 
Let us now return  to the equations of motion \eqref{neweuler}. If we consider an explicit model where $\Lambda = \Lambda(n^2 , \mathcal N^2)$, we have
\beq
{\partial \Lambda \over \partial \omega_{ab} } =  {\partial \Lambda \over \partial \mathcal N^2}  {\partial \mathcal N^2 \over\partial \omega_{ab} }=  {\partial \Lambda \over \partial \mathcal N^2} \omega^{ab}  = -{1\over 2} \lambda^{ab}\ ,
\eeq
and we arrive at 
\beq
n^a\omega_{ab} =  - {2 \over \kappa^2} \omega_{ab} \nabla_c \left(  {\partial \Lambda \over \partial \mathcal N^2} \omega^{ca}\right) = - {1 \over \kappa} \omega_{ab} \nabla_c \left(  {\partial \Lambda \over \partial \mathcal N} {1\over \mathcal N \kappa} \omega^{ca}\right) \ .
\label{final}\eeq

Making use of \eqref{omnew} we have
\beq
 {1 \over \kappa} \omega_{ab} \nabla_c \left(  {\partial \Lambda \over \partial \mathcal N} {1\over \mathcal N \kappa} \omega^{ca}\right)   = -  \mathcal N \perp^a_b \left[  \nabla_a \left( {\partial \Lambda \over \partial \mathcal N} \right) -  {\partial \Lambda \over \partial \mathcal N}  \left( \hat \kappa^c \nabla_c \hat \kappa_a  -  u^c \nabla_c u_a \right)\right] \ .
 \label{fvort}
\eeq
Here it is worth noting that $-\partial \Lambda /\partial \mathcal N$ is naturally interpreted as the energy per vortex (assuming that all vortices carry the same circulation and that the averaged energy is simply proportional to the vortex density.  It is straightforward to make a connection  with the ``thin vortex'' limit  considered in \cite{kr3} but we will not do so here.

Suppose that we also introduce a (distinct) four-velocity associated with the matter flux (the condensate), i.e. let
\beq
n^a = n u_\n^a \ ,
\eeq
such that 
\beq
u_\n^a = \gamma ( u^a + v^a) \ , \quad u^a v_a = 0 \ , \quad \gamma = (1-v^2)^{-1/2} \ .
\eeq
We then have 
\beq
n^a \omega_{ab} =n  \gamma \mathcal N v^a \kappa^d \epsilon_{dab}   =  n \gamma \mathcal N \epsilon_{bac} \kappa^a v^c \ .
\eeq
This represents the  Magnus force  that acts on a set of  vortices moving relative to a superfluid condensate (represented by $n^a$) (see for example cite{trev}. Also recognizing the tension associated with the bending of the vortices, we have the final equations of motion
\beq 
 {n \gamma  \epsilon_{bac} \kappa^a v^c}
  =    \perp^a_b \left[  \nabla_a \left( {\partial \Lambda \over \partial \mathcal N} \right) - { {\partial \Lambda \over \partial \mathcal N}  \hat \kappa^c \nabla_c \hat \kappa_a } +  {\partial \Lambda \over \partial \mathcal N}  u^c \nabla_c u_a \right] \ .
  \label{vomo}
\eeq

For completeness, and immediate benefit for the discussion of elasticity,  we should also work out stress-energy tensor for this model. This is fairly straightforward, as 
 the required calculation repeats \eqref{varlam}, apart from that we now need to account for the $\mathcal N^2$ dependence on
 $\Lambda$.  With $\Lambda = \Lambda(n^2, \mathcal N^2) = \Lambda(n_{abc}, \omega_{ab}, g^{ab})$ we need 
\beq
{\partial \Lambda \over \partial \mathcal N^2 } \delta \mathcal N^2  ={1\over 2 \mathcal N \kappa^2 } {\partial \Lambda \over \partial \mathcal N}   \left( g^{cd} \omega_{ca}   \omega_{db} \delta g^{ab}  +   \omega^{bd} \delta \omega_{bd} \right) \ ,
\eeq
leading to a contribution  (using \eqref{omnew})
\beq
{\partial \Lambda \over \partial \mathcal N^2 } {\delta \mathcal N^2 \over \delta g^{ab}} =  {1\over 2}   \mathcal N   {\partial \Lambda \over \partial \mathcal N} \perp_{ab}\ .
\eeq
Combining this with the previous (fluid) result, we have
\beq
T_{ab} = \left( \Lambda - n^c \mu_c \right) g_{ab} + n_a \mu_b -  \mathcal N   {\partial \Lambda \over \partial \mathcal N} \perp_{ab} \ .
\label{stressvort}
\eeq
A direct calculation verifies that the divergence of this expression leads us back to \eqref{vomo}. It is a straigtforward exercise, which involves \eqref{divk},  \eqref{divu}, \eqref{uNperp} and \eqref{kNperp}, 

\section{Two-dimensional elasticity}

The role of the two-dimensional vortex space should be clear from the derivation of \eqref{vomo}. The developments relied heavily on results that are easy to obtain once we introduce the $\chi^I$ coordinates. The power of this approach becomes even more apparent when we consider elastic aspects of the vortex lattice. This should be expected from the corresponding problem of the neutron star crust---where the geometry of the configuration space associated with the $X^A$ coordinates plays a central role \cite{KS,lagrange}. Adapting this argument to the two-dimensional case, we 
can account for 
 stresses and strains of the vortex lattice.

We focus our attention on two new matter-space tensors. First of all, we introduce another object that remains fixed along the flow; 
\beq
k_{ab} = \bar \psi^I_a \bar \psi^J_b k_{IJ}\ ,
\eeq
such that 
\beq
u^a k_{ab} = 0 \ ,
\eeq
and 
\beq
\mathcal L_u k_{ab} = 0 \ ,
\eeq
In essence, it  follows (as in the case of $\omega_{ab}$ that 
\beq
\Delta k_{ab} = 0 \ .
\eeq

Next, we introduce 
\beq
\eta_{ab} = \bar \psi^I_a \bar \psi^J_b \eta_{IJ}\ ,
\eeq
to represent the relaxed lattice configuration. This simply means that, in absence of elastic stresses we have \cite{KS,lagrange}
\beq
g^{IJ} \eta_{JK} = \delta^J_K\ ,
\eeq
where
\beq
g^{IJ} =  \bar \psi^I_a \bar \psi^J_b g^{ab} = \bar \psi^I_a \bar \psi^J_b \perp^{ab}\ .
\eeq
As the spacetime evolves with the system, we can represent the elastic strain associated with any deformation by
\beq
s_{AB} = {1\over 2} \left( g_{IJ} - \eta_{IJ} \right) \quad \Longrightarrow \quad s_{ab} = {1\over 2} \left( \perp_{ab} - \eta_{ab}\right)\ .
\eeq 

Finally, we focus on conformal deformations, for which (adapting the argument from the Appendix in \cite{lagrange} to the present two-dimensional setting) we have
\beq
k_{ab} = W \eta_{ab} = \mathcal N \kappa \eta_{ab}\ ,
\eeq
and---as we are mainly interested in modest effects---we also make the Hookean approximation
\beq
\Lambda = \check \Lambda (n^2, \mathcal N^2) - \check \mu(\mathcal N) s^2\ ,
\label{ellam}
\eeq
where the sign is motivated by the fact that the energy measured by an observer moving along with the vortex array (with four velocity $u^a$) is $\varepsilon = - \Lambda$ and $\check \mu$ represents the shear modulus. As in \cite{lagrange} we use checks to indicate that quantities are evaluated for the unstrained configuration. In effect, the first term in \eqref{ellam} remains as in the previous section, so we may focus on the second contribution. Clearly, this leads to
\beq
\delta \Lambda = \delta \check \Lambda - {\partial \check \mu \over \partial \mathcal N} \delta \mathcal N - \check \mu \delta s^2 \ .
\eeq
The middle term is readily evaluated using results we already have at hand. The final term is different, as we have to provide a form for the strain scalar $s^2$ in order to make progress. 

In general, we will have $s^2 = s^2(\mathcal N, k_{ab}, g^{ab})$, which means that
\beq
\delta \Lambda =  {\partial \check \Lambda \over \partial n^2} \delta n^2 + \left[ {\partial \check \Lambda \over \partial \mathcal N} - {\partial \check \mu \over \partial \mathcal N}s^2  - \check \mu {\partial s^2 \over \partial \mathcal N}  \right] \delta \mathcal N
 - \check \mu \left[ {\partial s^2 \over \partial k_{ab} } \delta k_{ab} + {\partial  s^2  \over \partial g^{ab}} \delta g^{ab} \right] \ .
\eeq
Focussing on terms associated with the vortex elasticity, we need
\beq
\delta \mathcal N  = { 1\over 2\mathcal N \kappa^2}  \left(   g^{ac} g^{bd}\omega_{cd}  \delta \omega_{ab}+ g^{cd} \omega_{ac} \omega_{bd} \delta g^{ab}   \right) \ ,
\eeq
where $\Delta \omega_{ab}=0$ allows us to shift the focus onto a variation with respect to the displacement $\xi^a$ (as before):
\beq
\delta \omega_{ab} = 0 \Longrightarrow \delta \omega_{ab} = - \xi^c \nabla_c \omega_{ab} - \omega_{cb} \nabla_a \xi^c - \omega_{ac} \nabla_b \xi^c\ .
\eeq
Similarly, we have
\beq
\Delta k_{ab} = 0 \Longrightarrow \delta k_{ab} = - \xi^c \nabla_c k_{ab} - k_{cb} \nabla_a \xi^c - k_{ac} \nabla_b \xi^c \ ,
\eeq
and it follows that the elastic contributions to the stress-energy tensor are (with the $\ldots$ representing terms that remain as in \eqref{stressvort})
\begin{multline}
{\delta \Lambda \over \delta g^{ab}} =  \ldots -   { \mathcal N \over 2}  \left[ {\partial \check \mu \over \partial \mathcal N}s^2  + \check \mu {\partial s^2 \over \partial \mathcal N}  \right] \perp_{ab} - \check \mu  {\partial  s^2  \over \partial g^{ab}}  \\
=   \ldots -   { \mathcal N \over 2}  {\partial \check \mu \over \partial \mathcal N}s^2  - \check \mu \left[  { \mathcal N \over 2} {\partial s^2 \over \partial \mathcal N} \perp_{ab} +  {\partial  s^2  \over \partial g^{ab}}   \right] \ .
\end{multline}

This may be as far as we can get without specifying the strain scalar $s^2$. 
An intuitive approach to that part of the problem \cite{KS,lagrange} is to build $s^2$ out of ``invariants'' of $\eta_{ab}$. In two dimensions it makes sense to use
\begin{equation}
I_1 = {\eta}^I_{\ I} = {1\over \mathcal N \kappa} g^{IJ} k_{IJ} \ ,
\eeq
and
\beq  
 I_2 = {\eta}^I_{\ J} {\eta}^J_{\ I} = {1\over (\mathcal N \kappa)^2} g^{IJ} g^{KL} k_{JK} k_{LI} \ .
\end{equation}
We then have
\beq
{\partial s^2 \over \partial \mathcal N}  = {\partial s^2 \over \partial I_1} {\partial I_1 \over \partial \mathcal N} +  {\partial s^2 \over \partial I_2} {\partial I_2 \over \partial \mathcal N}
= -{\partial s^2 \over \partial I_1} {I_1 \over \mathcal N} - 2  {\partial s^2 \over \partial I_1} {I_2 \over \mathcal N}  \ ,
 \eeq
 and
 \begin{multline}
{\partial s^2 \over \partial g^{ab}}  =  {\partial s^2 \over \partial I_1} {\partial I_1 \over \partial g^{ab}} +  {\partial s^2 \over \partial I_2} {\partial I_2 \over \partial g^{ab}} 
= \bar \psi^I_a \bar \psi^J_b \left[  {\partial s^2 \over \partial I_1} {\partial I_1 \over \partial g^{IJ}} +  {\partial s^2 \over \partial I_2} {\partial I_2 \over \partial g^{IJ}} \right] \\
 =  \bar \psi^I_a \bar \psi^J_b \left[  {\partial s^2 \over \partial I_1}  \eta_{IJ} +  2  {\partial s^2 \over \partial I_2}  g^{KL} \eta_{JK} \eta_{LI}  \right] \ ,
\end{multline}
which leads to
\beq
 { \mathcal N \over 2} {\partial s^2 \over \partial \mathcal N} \perp_{ab} +  {\partial  s^2  \over \partial g^{ab}}  
 =     {\partial s^2 \over \partial I_1}  \eta_{\langle ab \rangle}   + 2   {\partial s^2 \over \partial I_2}  \eta_{c\langle a} \eta_{b \rangle }^{\ c}  \ ,
 \eeq
 where the $\langle\ldots\rangle$ indicate that the trace is removed. The fact that this object is trace-free indicates that it represents anisotropic stresses. The result is (naturally) similar to that from the three-dimensional problem (see, e.g. equation (80) in \cite{lagrange}).
 
Putting the pieces together, the complete stress-energy tensor takes the form
 \begin{multline}
T_{ab} = \left( \check \Lambda - n^c \mu_c  - \check \mu s^2 \right) g_{ab} + n_a \mu_b -  \mathcal N  \left(  {\partial \check \Lambda \over \partial \mathcal N} - {\partial \check \mu \over \partial \mathcal N}  s^2 \right)\perp_{ab}  \\
+ 2 \check \mu \left(   {\partial s^2 \over \partial I_1}  \eta_{\langle ab \rangle}   + 2   {\partial s^2 \over \partial I_2}  \eta_{c\langle a} \eta_{b \rangle }^{\ c}   \right) \ .
\end{multline}
In order to work out the new terms in the equations of motions, we first write the elastic contributions
\begin{multline}
T_{ab} = \ldots - \check \mu s^2 g_{ab} +  \mathcal N  {\partial \check \mu \over \partial \mathcal N}  s^2 \perp_{ab}  
+ 2 \check \mu \left(   {\partial s^2 \over \partial I_1}  \eta_{\langle ab \rangle}   + 2   {\partial s^2 \over \partial I_2}  \eta_{c\langle a} \eta_{b \rangle }^{\ c}   \right) \\
= \ldots - \check \mu s^2 g_{ab} +  \mathcal N  {\partial \check \mu \over \partial \mathcal N}  s^2 \perp_{ab}  
+\pi_{ab} \ .
\end{multline}
That is,  we need
\begin{multline}
\nabla_a \left[  - \check \mu s^2 \delta^a_b +  \mathcal N  {\partial \check \mu \over \partial \mathcal N}  s^2 \perp^{a}_b   \right] \\
 = - s^2 {\partial \check \mu \over \partial \mathcal N} \nabla_b \mathcal N - \check \mu \nabla_b s^2 
 +  {\partial \check \mu \over \partial \mathcal N}   \left[ s^2\nabla_ b \mathcal N
+ \mathcal N  s^2 \nabla_a  \perp^{ab} +  \mathcal N   \nabla_ b s^2 \right] \ ,
\end{multline}
where we have used \eqref{Nperp2} and an analogous argument for $s^2$, which is also a matter space object.
It follows that the elastic contribution is
\beq
\nabla_a T^a_{\ b} =  \ldots+  \perp^a_b \left[   \left( \mathcal N {\partial \check \mu \over \partial \mathcal N}  - \check \mu \right)  \nabla_a s^2  + \mathcal N   {\partial \check \mu \over \partial \mathcal N}  s^2 \left( u^c \nabla_c u_a - \kappa^c \nabla_c \kappa_a  \right) \right]  + \nabla_a \pi^a_{\ b} \ ,
\eeq
and the complete equations of motion take the form
\begin{multline}
n^a \omega_{ab} \\
 =  \perp^a_b \left[ \nabla_a \left(  {\partial \Lambda \over \partial \mathcal N}  \right) -    \left( \mathcal N {\partial \check \mu \over \partial \mathcal N}  - \check \mu \right)  \nabla_a s^2 
  +\left(  \mathcal N   {\partial \Lambda \over \partial \mathcal N} -  \mathcal N   {\partial \check \mu \over \partial \mathcal N}  s^2\right) \left( u^c  \nabla_c u_a -  \hat \kappa^c  \nabla_c  \hat \kappa_a \right)  \right] \\
  - \nabla_a \pi^a_{\ b} \ .
  \label{finaleq}
\end{multline}
This is the final result, describing the dynamics of an elastic vortex array in full general relativity. It can be meaningfully compared to the corresponding relation for an elastic nuclear lattice, e.g. equation~(85) in \cite{lagrange}. Notably, the model is nonlinear (although the Hookean assumption \eqref{ellam} obviously implies a Taylor expansion for weak strains). This is in contrast to previous (Newtonian) models, which have exclusively been perturbative.

\section{Perturbations and the Newtonian limit}

If we want to compare the final equations of motion \eqref{finaleq} to the corresponding expression in the Newtonian context  \cite{baymc,cbaym,baym} we need to do a bit more work. The usual expression for vortex elasticity tends to be given in terms of displacement vectors, i.e. at the perturbative level. In order to facilitate a comparison we need to reframe the result in terms of explicit (Lagrangian) perturbations with respect to an unstrained background configuration. The strategy for doing this is the same as in the case of an elastic nuclear lattice, e.g. the problem considered in \cite{lagrange}. The fact that the elastic contribution is two-dimensional makes little conceptual difference. However, we need to pay careful attention to the unperturbed configuration. 

A suitable background configuration involves two spacetime symmetries. First of all, the assumption that the problem is stationary implies the existence of a timelike Killing vector. Taking the four velocity of the configuration to the aligned with this Killing vector, it follows immediately that 
\beq
u^a \nabla_a u_b = 0 \ .
\eeq
This is natural and intuitive---there is no acceleration associated with the equlibrium configuration. We also need to consider the vortex array, which in equilibrium ought to be associated with axisymmetry. Letting the vortex vector $\hat \kappa^a$ be aligned with a second (spatial) Killing vector, we see that we should also have
\beq
\hat \kappa^a \nabla_a \hat \kappa_b = 0 \ .
\eeq
The implications are that there is no contribution from the tension---intuitively, the vortices are ``straight''---and the vortex array moves without expansion/contraction. Again, this makes sense. Finally, the absence of elastic strain implies that background is such that both $s^2$ and $\pi^a_{\ b}$ vanish. In practice, we have 
  \beq
  \eta_{ab} = \perp_{ab} \ ,
  \eeq
for a relaxed configuration.

Turning to the perturbed case and the anisotropic stresses, it  is helpful to make the model (even more) specific. 
In order to construct a suitable combination to represent the strain scalar, we first of all note that $I_1 = I_2=2$ in the relaxed configuration. Secondly, we know that $k_{IJ}$ is (by construction) independent of $W$ (or equivalently $\mathcal N$) so we can scale out the dependence on this quantity if we work with $I_1^2$ and $I_2$. A simple possibility would then be 
\beq
s^2 = I_2 - {1\over 2} I_1^2\ .
\label{s2ex}
\eeq
Other prescriptions are, of course, available, but we will take \eqref{s2ex} as our example. We then have
\beq
{\partial s^2 \over \partial I_1} = - I_1 \quad \text{and} \quad {\partial s^2 \over \partial I_2} = 1 \ .
\eeq
This means that 
\begin{multline}
 \pi^a_{\ b} =  -  2 \check \mu g^{ad} \left(   I_1 \eta_{\langle db \rangle}   - 2   \eta_{c\langle d} \eta_{b \rangle }^{\ c}  \right) \\
 = -  2 \check \mu g^{ad} \left[   I_1 \left( \eta_{db} - {1\over 2} I_1 g_{db }\right)   - 2  \left( \eta_{cd} \eta_{b}^{\ c} - {1\over 2} I_2  g_{db}\right) \right] \ .
 \end{multline}
 As the combined terms in the bracket vanish for an unstrained background, which we perturb with respect to, we have
 \beq
 \Delta \pi^a_{b} = -  2 \check \mu g^{ad} \Delta  \left[   I_1 \left( \eta_{db} - {1\over 2} I_1 g_{db }\right)   - 2  \left( \eta_{cd} \eta_{b}^{\ c} - {1\over 2} I_2  g_{db}\right) \right] \ .
  \eeq
  We now need
  \beq
  \Delta \mathcal N 
 =  {\mathcal N\over 2} \bar \psi^I_a \bar\psi^J_c  g_{IJ}\Delta g^{ac}
 = {1\over 2} \mathcal N \perp_{ac} \Delta g^{ac} \ .
  \eeq
  Combining this with $\Delta k_{ab}$, we have
  \beq
\Delta \eta_{ab} = \Delta \left(  {1\over \mathcal N \kappa} k_{ab}  \right) =-  {1\over \mathcal N^2 \kappa} k_{ab} \Delta \mathcal N
= - {1 \over 2} \eta_{ab} \perp_{cd} \Delta g^{cd} \ .
  \eeq
    It also follows that 
  \beq
  \Delta I_1 = \Delta \left( g^{ab} \eta_{ab} \right) =  \perp_{ab} \left( \Delta g^{ab}  - {1 \over 2} g^{ab}  \perp_{cd} \Delta g^{cd} \right)  = 0 \ ,
  \eeq
  since $g^{ab} \perp_{ab} = \delta^a_a -1-1 = 2$, and 
    \beq
  \Delta I_2 = \Delta \left( g^{ac} g^{bd} \eta_{ab} \eta_{cd} \right) 
  = 2 \perp_{ac} \Delta g^{ac} +  2 \perp^{cd} \left(-{1\over 2} \perp_{cd} \perp_{ab} \Delta g^{ab} \right)  = 0 \ .
  \eeq
   Putting things together, we are left with
  \beq
    \Delta \pi^a_{\ b} 
  =   4 \check \mu g^{ad} \left[  \perp_{fd} \perp_{be}  -{1 \over 2} \perp_{db} \perp_{ef} \right]  \Delta g^{ef} \ .
  \eeq
This results is notably similar to the corresponding expression of an elastic nuclear lattice, e.g. equation~(91) in \cite{lagrange}. 
The main difference is that the projection is no orthogonal to both $u^a$ and $\hat \kappa^a$ and the two-dimensional nature of the lattice leads to a factor of $1/2$ rather than the usual $1/3$ (this is simply the inverse of the number of dimensions of the elastic matter). In essence, the result makes intuitive sense.

  The final step involves expressing the contribution to the perturbed equations of motion in terms of the displacement vector. In order to do this, we need
 \beq
 \Delta \nabla_a \pi^a_{\ b} = \nabla_a  \Delta \pi^a_{\ b} \ ,
 \eeq
 which is true as long as the background is unstrained.  This then means that 
 \beq
 \Delta \pi^a_{\ b} =   4 \check \mu  \left[  \perp^{a}_{\ f} \perp_{be}  -{1 \over 2} \perp^a_{\ b} \perp_{ef} \right]  \delta g^{ef}  
+    4 \check \mu  \left( \perp^a_{\ b} \perp^c_{\ e} -  \perp^{ac}  \perp_{be} -   \perp^a_{\ e} \perp^c_{\ b} \right)  \nabla_c \xi^e \ .
  \eeq
  At this point it is natural to introduce the totally projected derivative (with a projection of each  free index)
  \beq
  D_i \xi^j = \perp_i^a \perp^j_b \nabla_a \xi^b \ .
  \eeq
  This is helpful as it facilitates an immediate comparison with Newtonian expressions. 
 We then have
  \begin{multline}
 4 \check \mu  \left( \perp^a_{\ b} \perp^c_{\ e} -  \perp^{ac}  \perp_{be} -   \perp^a_{\ e} \perp^c_{\ b} \right)  \nabla_c \xi^e \\
      = - 4 \check \mu\perp^{ai} \perp_b^{\ j} \left[  D_i \xi_j +  D_j \xi_i  - g_{ij}  D_l \xi^l  \right] 
   = - \perp^{ai} \perp_b^{\ j} \Pi_{ij}\ ,
    \end{multline}
    where we have defined
    \beq
    \Pi_{ij} = 4 \check \mu \left[  D_i \xi_j +  D_j \xi_i  - \perp_{ij}  D_l \xi^l  \right] \ ,
    \eeq
   which  is manifestly orthogonal to both $u^a$ and $\hat \kappa^a$. This leads to
       \beq
    \nabla_ a \left(  \perp^{ai} \perp_b^{\ j} \Pi_{ij} \right) 
    = D^l \Pi_{lb} +  \Pi_{ib}  \left( u^a \nabla_a u^i - \hat \kappa^a \nabla_a \hat \kappa^i \right) 
 +  \Pi^a_{\ j}  \left( u_b \nabla_a u^j - \hat \kappa_b  \nabla_a \hat \kappa^j \right)\ .
 \label{long}
    \eeq
 We have already argued that the second term on the right-hand side should vanish for a suitable equillibrium configuration, and it is easy to show (since $\Pi^a_{\ b}$ is symmetric) that the Killing vector argument removes the third term, as well. We are left with
      \beq
    \nabla_ a \left(  \perp^{ai} \perp_b^{\ j} \Pi_{ij} \right) 
    = D^i \Pi_{ib}  = 4 D^i  \left[  \check \mu \left( D_i \xi_b +  D_b \xi_i  - \perp_{ib}  D_l \xi^l  \right) \right] \ .
    \eeq
This result resembles the result from the Newtonian setting, e.g. equation (24) in \cite{cbaym}, although one has to keep in mind that the derivatives do not commute in a curved spacetime. 

\section{Final remarks}

Starting from the well-established variational approach to relativistic fluids, and bringing in notions from a basic description of string dynamics, we have developed a model for superfluid dynamics. This provides a valuable---if somewhat technical---extension to previous efforts to account for elastic properties of a vortex array.
The approach to the problem is of conceptual interest as it highlights the role of the two-dimensional subspace orthogonal to a given vortex array (analogous to the world sheet coordinates used to describe a moving string in spacetime). The discussion also provides a detailed description of concepts that have previously been described in a somewhat phenomenological manner, and which may be applied to interesting problems in astrophysics and cosmology.  

 \acknowledgements

NA gratefully acknowledges helpful conversations with James Vickers and financial support  from STFC via grant ST/R00045X/1.

\end{document}